\documentclass[seceq]{ptptex}
\usepackage{graphicx} 
\usepackage{latexsym}
\notypesetlogo
\begin{document}
%
\vspace{10mm}
\begin{center}
\Large{ \bf Schwinger-Dyson equation on the complex plane }

\vspace{5mm}

\Large{\bf $-$ A four-fermion interaction model at finite temperature $-$}

\vspace{15mm}
   
\large{ Hidekazu {\sc Tanaka} \footnote{E-mail:tanakah@rikkyo.ac.jp} and Shuji {\sc Sasagawa} \\
Rikkyo University, Tokyo 171-8501, Japan\\
}
 \end{center}

\begin{center}

\vspace{25mm}

{\Large ABSTRACT}
 \end{center}
        
\vspace{10mm}

We extend the Schwinger-Dyson equation (SDE)  on the complex plane, which was treated in our previous research, to finite temperature. 
 As a simple example, we solve the SDE for a model with four-fermion interactions in the (1+1) space-time dimensions at strong coupling region. We investigate the properties of the effective mass and energy for the fermions, especially near the phase transition temperature.
 
\def\proj{{\bf P}}
\def\slsh#1{{#1}{\kern-6pt}/{\kern1pt}}



      



\newpage

\section{Introduction}

In quantum field theory, the Schwinger-Dyson equation (SDE) [1,2]  is useful as a method for performing non-perturbative calculations for strongly coupled systems, such as quantum chromodynamics (QCD) in the infrared region.(See reviews in Refs [3-5].) 
 In many studies, the SDEs are computed in Euclidean momentum space.
 However, the SDE has been extended to Minkowski momentum space as well.
  One of the motivations for extending the SDE to Minkowski momentum space is to investigate the behavior of the propagator in the time-like region.
   Even in the perturbative region, the propagator has poles and branch-cut along the real axis in the time-like four-momentum.

  The properties of the propagator in the non-perturbative region are even more non-trivial. 
  For example, the behavior of the  gluon propagator in the infrared region has been of interest. 
  Although calculations by lattice simulations are performed in Euclidean space, the numerical continuation of the data into Minkowski space suggests the existence of poles in the complex squared momentum plane in the infrared region [6-8]
   and the violation of positive values of the spectral functions [9-11], which should have positive values if the asymptotic fields exist.  
  These results were compared with predictions from the Gribov-Zwanziger approach [12-17] and a calculation by the SDE [18].
   
One of the computational difficulties using the SDE in Minkowski space is the presence of poles in the propagator.
 In the SDE, the effective mass is calculated using the self-energy,
 which requires knowledge of the precise pole positions of the propagator.  
To avoid this problem, the Wick rotation from the real axis to the imaginary axis (or opposite direction) is often used in the momentum integration. However, the Wick rotation requires the location of the poles to be known in advance, but the value of an effective mass in  non-perturbative region is non-trivial in calculation with the SDE. 
 \footnote{Some formal approaches have been done in Refs.[19,20],
  in which the gluon propagator with complex singularities in Minkowski space is reconstructed starting from the Euclidean propagator by the analytic continuation. } 
   
 The analytical properties of the gluon propagator are studied by connecting the square of momentum from space-like to time-like momentum, or by extending it to complex values.   
   These studies suggest that the gluon propagator has complex poles in the non-perturbative region [21,22].
 
 It has also been pointed out that the fermion (electron or quarks) propagator has complex poles in the strong coupling region.
 In investigation using the SDE, the effective propagators for the fermions in the strong coupling region are calculated using the one-loop self-energy with the effective propagator itself [23-27] or the spectral representation [28-33] in the integrand, in which the square of momentum changes from a positive value to a negative value, or extends it to a complex value.
 
 However, recent analysis of data from lattice simulations using the Pad\'e approximation has not confirmed the existence of the complex poles for the quark propagator[34].
 On the other hand, using the SDE, Ref.[35] showed the existence of a pair of complex conjugate poles located very close to the real axis.

Considering these situations, it seems meaningful to investigate the analytical properties for the propagator of the fermions in the non-perturbative region using various methods. 
 From another point of view  as a more general problem setting for the SDE, it seems interesting to set up a problem of what kind of solutions can be obtained by the SDE for various models with complex structures in the strongly coupled region. This may give some hints or new insights when considering the above problem and other applications. Another interesting issue is the behavior of the propagator near phase transition points at finite temperature.
In these studies, the properties of the spectral functions were mainly investigated [36-39].

  In our previous paper [40], we used the SDE defined on the complex energy plane to investigate the positions of the poles of the effective propagator at zero temperature.  Our method  differs  slightly from other methods by defining the SDE as having complex mass and complex energy from the beginning without assuming a spectral representation, and the self-energy of the fermion is integrated  with  arbitrary integral paths on the complex energy plane. 
 Then, we solved for the real and imaginary parts of the SDE simultaneously.

 We explained our formulation and analysis method using two simple examples.
First, in order to illustrate our method, we examined the SDE for a contact four-fermion interaction model in the $(1+1)$ space-time dimensions, such as   Gross-Neveu (GN) model for chiral fermions at large flavor limit [41.42] with the effective mass which does not depend on the energy and momentum.
We found that the poles of the effective propagator lie on the real axis of the complex energy plane, even though the effective mass has an imaginary part. 

 We also found that the effective mass obtained by the SDE depends on the   respective signs of the initial input   values in the iterative method.   From our calculations, the SDE may lead to multiple solutions depending on the initial input values.
Moreover, for the integral path including the real axis (Minkowski momentum integration), the solutions obtained by the iterative method  oscillate between solutions obtained by the integral path including the imaginary axis (Euclidean momentum integration). On the other hand, the energy does not depend on these two integral paths.

As a second example, we investigated the strongly coupled QED with the instantaneous exchange approximation [43,44]. In this model, the mass function depends on the spatial momentum.   We found that the imaginary part of the energy is positive valued over the defined momentum range when the real part of the energy has positive value. As discussed in the previous paper, the effective propagator behaves anomalously in non-perturbative region.  

It may be interesting to investigate how the properties obtained at zero temperature in the previous studies change at finite temperature.
  For analyzes at a finite temperature, it is more appropriate to use a complex energy rather than the complex square of the momentum. 
Before applying our method to gauge theories,  we examine a simple model used in the previous work.

In this paper, we extend the SDE on the complex energy plane to finite temperature. In our method, the mass and energy of the SDE are extended to the complex values, and the propagator in the integrand of the self-energy is integrated over the complex energy plane as used in the previous paper.
    We explain our formulation and analysis method using a simple model with four-fermion interactions in the $(1+1)$ space-time dimensions.
 
Our paper is organized as follows:
In Section 2, we formulate the SDE for the effective mass  at finite temperature.  In Section 3, using the SDE, we numerically calculate the effective mass and energy.  Section 4 is devoted to a summary and some comments. Explicit formulas using the numerical calculations are shown in Appendix A. In Appendix B, we present alternative formulas for numerical calculations.

\section{The SDE for a four-fermion interaction model on the complex plane}

In this section, we investigate a simple SDE for the effective mass $M$ in order to illustrate our method.  \footnote{The GN model[41] was extended to finite temperature.[42,45-49].} We start from the SDE for the model we dealt with in the previous paper [40] at zero temperature as
\begin{eqnarray}
M=i\lambda \int dq_0dq{M \over q_0^2-q^2-M^2+i\varepsilon}
\end{eqnarray}
with a momentum $Q=(q_0,q)$ and $\lambda$ is a coupling constant.
 Here, we extend Eq.(2$\cdot$1) to finite temperature $T$ in the imaginary-time formalism, in which we replace $q_0$ to the Matsubara frequency $i\omega_n=2i\pi T(n+1/2) ~(n=0,\pm 1,\pm 2, \cdots$) for fermions and integration over  $q_0$ to $2\pi T i\sum_{n=-\infty}^{\infty} $, respectively.
Then, Eq.(2$\cdot$1) is given as 
\begin{eqnarray}
  M(T)= 2\pi T\lambda \int dq \sum_{n=-\infty}^{\infty} {M(T) \over  \omega_n^2+q^2+M^2(T)-i\varepsilon}\equiv \lambda \int dq M(T)I(T,q)
\end{eqnarray}
with
\begin{eqnarray}
 I(T,q)= 2\pi T \sum_{n=-\infty}^{\infty} {1 \over  \omega_n^2+q^2+M^2(T)-i\varepsilon}.
\end{eqnarray}
The sum of the discrete Matsubara frequency $\omega_n$ can be written by the integral form [50] as
\begin{eqnarray}
I(T,q)={1 \over 2}I^{(+)}(T,q)+{1 \over 2}I^{(-)}(T,q)
\end{eqnarray}
with
\begin{eqnarray}
I^{(+)}(T,q)=i \int_{-i\infty+\eta}^{i\infty+\eta} dz{1 \over z^2-q^2-M^2(T)+i\varepsilon}\zeta(T,z)
\end{eqnarray}
and
\begin{eqnarray}
I^{(-)}(T,q)=i \int^{-i\infty-\eta}_{i\infty-\eta} dz{1 \over z^2-q^2-M^2(T)+i\varepsilon}\zeta(T,z),
\end{eqnarray}
respectively, with a complex variable $z=z_{\rm R}+iz_{\rm I}$. 
 Here, we define
\begin{eqnarray}
 \zeta(T,z)=\tanh\left({1 \over 2}\beta z\right)
 \end{eqnarray}
with $\beta=1/T$. Since $\zeta(T,z)$ in $I^{(\pm)}(T,q)$ has poles at $z=i\omega_n~(n=0,\pm 1,\pm 2, \cdots$), we shift the integral paths by $\pm \eta$ from the imaginary axis and  finally we take $\eta \rightarrow +0$.

In order to evaluate the effective mass on the complex $z$ plain, we extend the effective mass $M(T)$ to a complex value, which is written as $M(T)=M_{\rm R}(T)+iM_{\rm I}(T)$. 

 In the following calculations, we write the denominator in the integrand  as 
\begin{eqnarray}
z^2-q^2-M^2(T)+i\varepsilon\equiv z^2-E^2(T,q)= (z-E(T,q))(z+E(T,q)),
\end{eqnarray}
where the complex energy $E(T,q)$ is defined by $E(T,q)\equiv E_{\rm R}(T,q)+iE_{\rm I}(T,q)$ with  $ E_{\rm R}(T,q)> 0$. \footnote{ Explicit relations among the complex values are given in Ref.[40].}
Therefore, the poles of the propagator are located at $z=\pm E(T,q)$ on the complex $z$ plane.

Here, we write Eqs.(2$\cdot$5) and (2$\cdot$6) as
\begin{eqnarray}
 I^{(\pm)}(T,q)=i\oint_{C_{\pm}} dz {1 \over z-z_{\pm}}f ^{(\pm)}(T,z,q)
\end{eqnarray}
and 
\begin{eqnarray}
 f^{(\pm)}(T,z,q)={1 \over z+z_{\pm}}\zeta(T,z) 
\end{eqnarray}
with $z_{\pm}=\pm E(T,q)$, in which we define $C_+$ as the integral path surrounding the pole $z=z_+$ in the right half-plane and $C_-$ as the integral path surrounding the pole $z=z_-$ in the left half-plane.
The explicit representations of the integral are given as  
$$I^{(+)}(T,q)=i\int_{-i\infty + \eta}^{i\infty + \eta} dz{1 \over z-z_+}f^{(+)}(T,z,q) $$
\begin{eqnarray}
+\lim_{\Lambda_0\rightarrow\infty}i\int_{\pi/2}^{-\pi/2} \Lambda_0e^{i\theta}id\theta {1 \over \Lambda_0e^{i\theta}-z_+}f^{(+)}(T,z=\Lambda_0e^{i\theta},q) 
\end{eqnarray}
and
$$I^{(-)}(T,q)= i\int_{i\infty - \eta}^{-i\infty - \eta} dz{1 \over z-z_-}f^{(-)}(T,z,q) $$
\begin{eqnarray}
+\lim_{\Lambda_0\rightarrow\infty}i\int^{\pi/2}_{3\pi/2} \Lambda_0e^{i\theta}id\theta {1 \over \Lambda_0e^{i\theta}-z_-}f^{(-)}(T,z=\Lambda_0e^{i\theta},q), 
\end{eqnarray}
respectively. Here, the $\theta$ integral for $\Lambda_0 \rightarrow \infty$  
vanishes as $f^{(\pm)}(T,z=\Lambda_0e^{i\theta},q)\rightarrow O(1/\Lambda_0)\rightarrow 0.$

By the closed path integrals along the integral paths $C_{\pm}$ on the complex $z$ plane, the residues of the two poles are obtained and $ I^{(\pm)}(T,q)$ are given by
\begin{eqnarray}
 I^{(\pm)}(T,q)=i\oint_{C_{\pm}} dz{1 \over z-z_{\pm}}f^{(\pm)}(T,z,q)
    = {\pi \over  E(T,q)} \zeta(T,E(T,q)),
\end{eqnarray}
 since  $z_{\pm}=\pm E(T,q)$ and $\zeta(T,-E(T,q))=- \zeta(T,E(T,q))$.

Finally, we obtain $ I(T,q)$ as
 \begin{eqnarray}
 I(T,q)={1 \over 2}I^{(+)}(T,q)+{1 \over 2}I^{(-)}(T,q)
={\pi \over  E(T,q)} \zeta(T,E(T,q))
\end{eqnarray}
 and the SDE for the effective mass is given by
\begin{eqnarray}
 M(T)= \pi \lambda\int dq { M(T) \over  E(T,q)} \zeta(T,E(T,q)).
\end{eqnarray}


\section{Numerical results}

In this section,  we numerically calculate the effective mass and energy using the SDE given in Eq. (2$\cdot$15).
 Introducing a ultraviolet cutoff $\Lambda$ and an infrared cutoff $\delta$ for the momentum $q$, we write  Eq. (2$\cdot$15) as
$$M(T)= \pi \lambda\int_{-\Lambda}^{\Lambda}  dq { M(T) \over  E(T,q)} \zeta(T,E(T,q)) \Theta(|q|-\delta)$$
\begin{eqnarray}
=2\pi \lambda\int_{\delta}^{\Lambda}  dq { M(T) \over  E(T,q)} \zeta(T,E(T,q)).
\end{eqnarray}
Here, $\Theta(|q|-\delta)$ denotes the Heaviside step function for restriction of the momentum $q$.

In numerical calculations, we write the SDEs for the real and imaginary parts of the effective mass as 
\begin{eqnarray}
M_{\rm R}(T) =2\pi\lambda\int_{\delta}^{\Lambda} dq {[M(T)E^*(T,q)\zeta(T,E(T,q))]_{\rm R} \over |E(T,q)|^2},  \end{eqnarray} 
and
\begin{eqnarray}
M_{\rm I}(T) =2\pi\lambda\int_{\delta}^{\Lambda} dq {[M(T)E^*(T,q)\zeta(T,E(T,q))]_{\rm I} \over |E(T,q)|^2},\end{eqnarray}
respectively, with $|E(T,q)|^2= E_{\rm R}^2(T,q)+ E_{\rm I}^2(T,q)$.
 Explicit expressions for $[M(T)E^*(T,q)\zeta(T,E(T,q))]_{\rm R} $  and $[M(T)E^*(T,q)\zeta(T,E(T,q))]_{\rm I} $ are given in Appendix A.

We solve the SDEs by iterative method from some initial input values for the real and imaginary parts of the effective  mass denoted by $M_{\rm R(0)} $ and $M_{\rm I(0)} $. In the following calculations, we set initial input values for the  effective   mass as $ M_{\rm R(0)}= M_{\rm I(0)}=0.01\Lambda$ \footnote{The signs of $E_{\rm R}(T,q)$ and $E_{\rm I}(T,q)$ are independent of the signs of the initial input values $M_{\rm R(0)}$ and $M_{\rm I(0)}$. However the solutions of  $M_{\rm R}(T)$ and $M_{\rm I}(T)$ depend on the sign of the initial input value $M_{\rm R(0)}$ as
$$M_{\rm R}(T)=s_{\rm R(0)}|M_{\rm R}(T)|,~~M_{\rm I}(T)=s_{\rm R(0)}|M_{\rm I}(T)|$$ with $s_{\rm R(0)}=M_{\rm R(0)}/|M_{\rm R(0)}|$. (See Ref.[40] at $T=0$.) In this paper, we show the solutions only for $ M_{\rm R(0)}>0$.} with $\delta/\Lambda=10^{-3}$ and $\varepsilon=10^{-5}\Lambda^2$ for $\lambda=0.3,0.6,0.9$, respectively.

Figs.1 and 2 show the temperature dependences of the real and imaginary parts of the effective mass divided by $\Lambda$, respectively.
 As shown in the Figures, the effective mass vanishes above a temperature $T_{\rm C}$, at which the chiral symmetry is restored. The phase transition temperature $T_{\rm C}$ depends on the coupling constant $\lambda$.
 In our numerical results, the phase transition temperatures divided by $\Lambda$ are given by $0.90<T_{\rm C}/\Lambda<0.93$ for $\lambda =0.3$, $1.86<T_{\rm C}/\Lambda<1.89$ for $\lambda =0.6$ and $2.79<T_{\rm C}/\Lambda<2.82$ for  $\lambda =0.9$, respectively.
\begin{figure}
\centerline{\includegraphics[width=8cm]{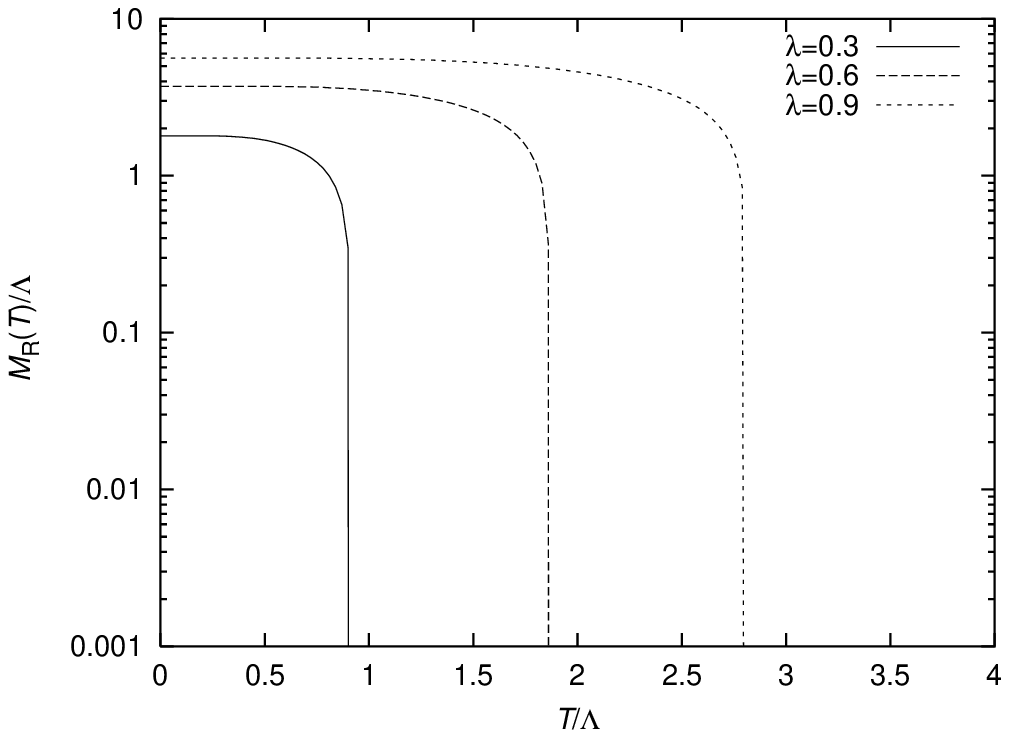}}
\caption{The temperature dependences of $M_{\rm R}(T)/\Lambda$ for $\lambda=0.3,0.6,0.9$. The horizontal axis denotes the temperature divided by  $\Lambda$.}
\end{figure}
 \begin{figure}
\centerline{\includegraphics[width=8cm]{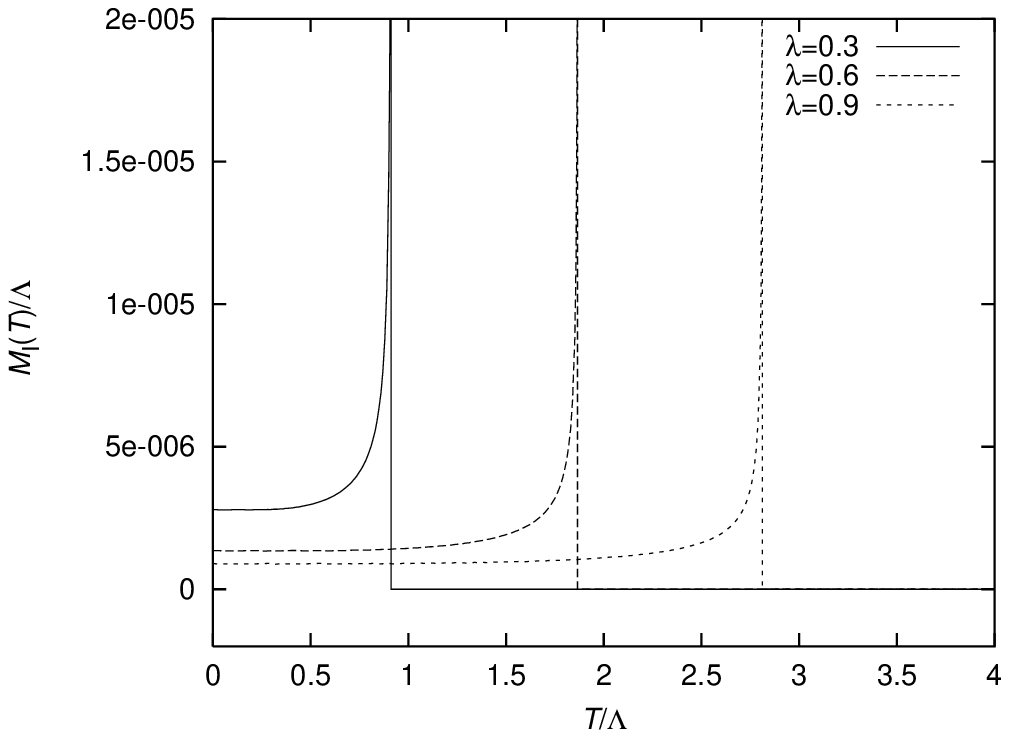}}
\caption{ The temperature dependences of $M_{\rm I}(T)/\Lambda$ for $\lambda=0.3,0.6,0.9$. The horizontal axis denotes the temperature divided by  $\Lambda$.}
\end{figure}
Fig.2 shows that the imaginary part of the effective mass increases at temperatures slightly below $T_{\rm C}$.\footnote{In our calculations, the imaginary part of the effective mass depends on the choice of $\varepsilon$, but the real part of the effective mass and the phase transition temperature are not sensitive to the value of $\varepsilon$.}  We will discuss this later.

In Appendix B, we present an alternative formulas which reproduce the numerical results presented in Figs.1 and 2. Moreover, the relation between the phase transition temperature $T_{\rm C}$ and the coupling constant $\lambda$ is approximately given by $T_{\rm C}/\Lambda\simeq \pi \lambda$, which is consistent with our numerical results.
\begin{figure}
\centerline{\includegraphics[width=8cm]{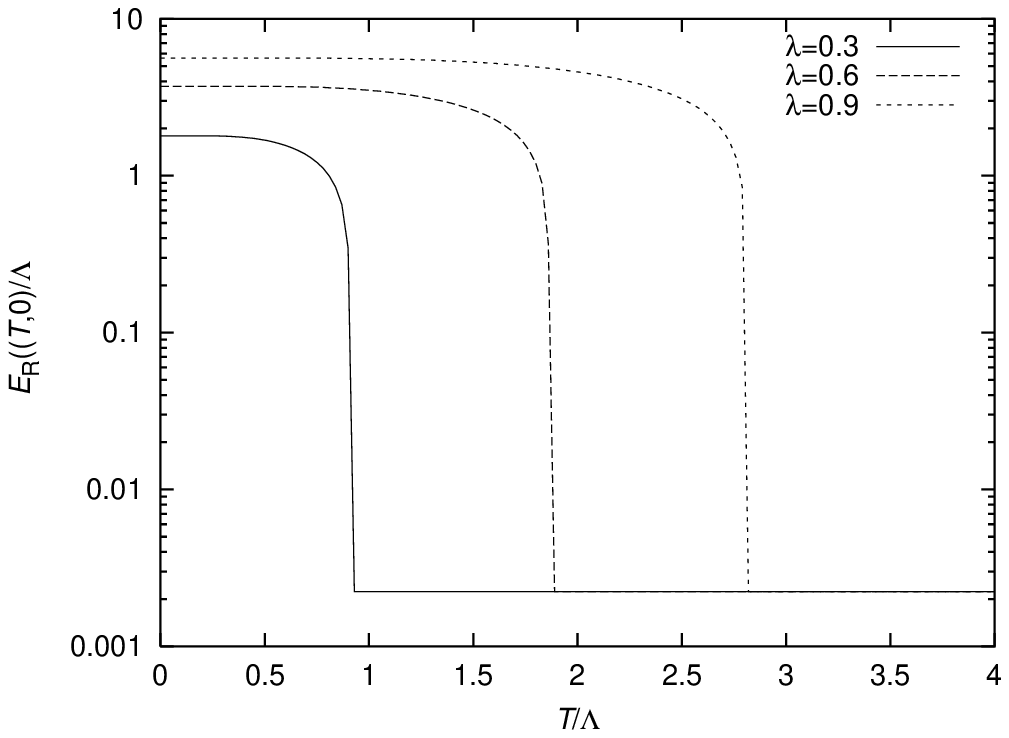}}
\caption{ The temperature dependences of $E_{\rm R}(T,q)/\Lambda$ with $q=0$ for $\lambda=0.3,0.6,0.9$. The horizontal axis denotes the temperature divided by  $\Lambda$.}
\end{figure}
\begin{figure}
\centerline{\includegraphics[width=8cm]{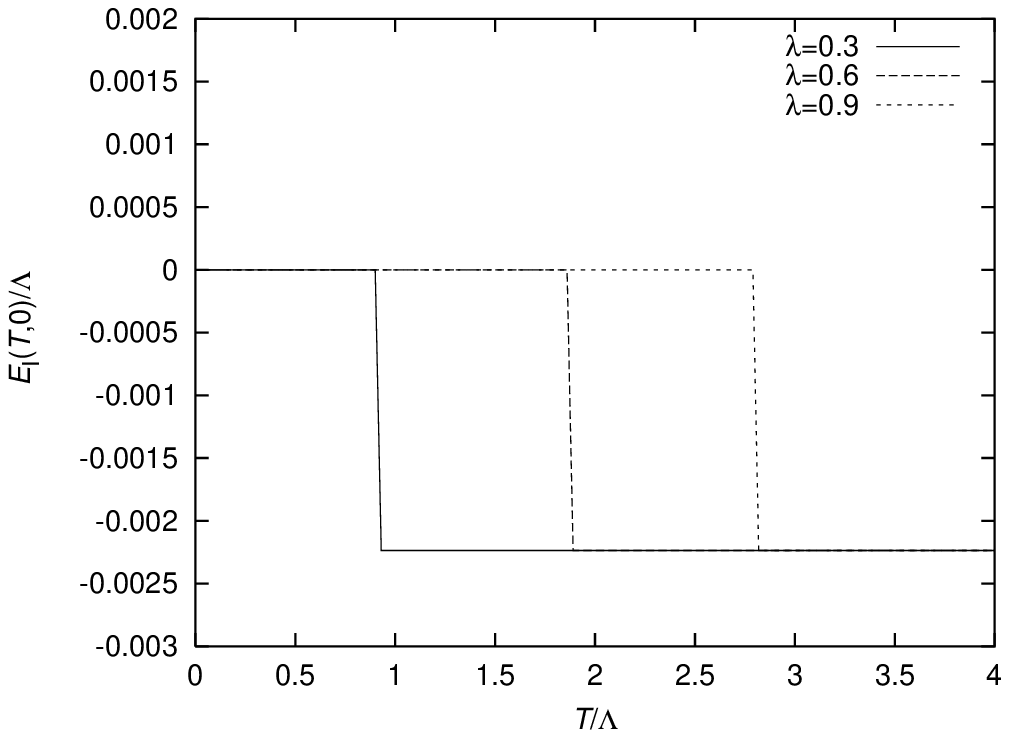}}
\caption{ The temperature dependences of $E_{\rm I}(T,q)/\Lambda$ with $q=0$ for $\lambda=0.3,0.6,0.9$. The horizontal axis denotes the temperature divided by  $\Lambda$.}
\end{figure} 

In Figs.3 and 4, we show the temperature dependences of the real and imaginary parts of the energy with the momentum $q=0$, respectively. 
 Since the real part of the energy is defined to be positive, the numerical results do not depend on the signs of initial input values of the effective mass.
The imaginary part of the energy converges as $E_{\rm I}(T,0)\rightarrow 0$ for $T<T_{\rm C}$ and it converges a constant value for $T>T_{\rm C}$. Moreover, the real part of the energy is also a constant value for $T>T_{\rm C}$.

As shown in Figs.1 to 4, around the phase transition temperature $T_{\rm C}$, the effective mass and energy change rapidly with temperature $T$. 
 In Fig.5, we show an enlarged view of the temperature dependence for the effective mass and energy around $T\simeq T_{\rm C}$ at $\lambda=0.3$. 
 The behaviors of the effective mass and energy  near $T_{\rm C}$ are similar for other coupling constants.

\begin{figure}
\centerline{\includegraphics[width=8cm]{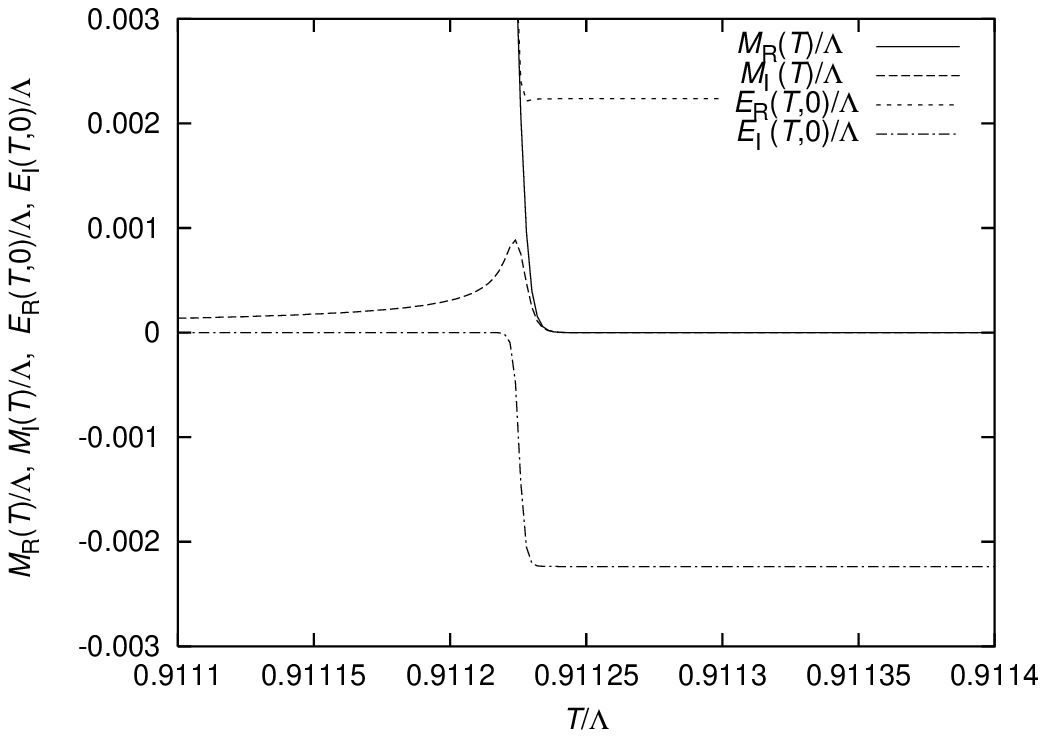}}
\caption{ The temperature dependences of $M_{\rm R}(T)/\Lambda$,$M_{\rm I}(T)/\Lambda$,$E_{\rm R}(T,q)/\Lambda$ and $E_{\rm I}(T,q)/\Lambda$ with $q=0$ for $\lambda=0.3$. The horizontal axis denotes the temperature divided by  $\Lambda$.}
\end{figure}

Below, we will separately consider the properties of the solutions obtained by the SDE in $T<T_{\rm C}$ and $T>T_{\rm C}$, respectively.

From Figs.4 and 5, except near $T_{\rm C}$, 
the imaginary part of the energy $E_{\rm I}(T,0)=0$ continues to hold for $0<T<T_{\rm C}$ as in the case of $T=0$ [40].
 In this region, we have \footnote{From Ref.[40], the real and imaginary parts of the energy are given by 
$$E_{\rm R}=\sqrt{(E^2)_{\rm R}+|E^2| \over 2},~~ E_{\rm I}={(E^2)_{\rm I}\over 2E_{\rm R}}$$
for $ E_{\rm R}>0$ with 
$ (E^2)_{\rm R}=q^2+(M^2)_{\rm R}=q^2+M_{\rm R}^2-M_{\rm I}^2$,
$(E^2)_{\rm I}=(M^2)_{\rm I}-\varepsilon= 2M_{\rm R}M_{\rm I}-\varepsilon $
and $|E^2|=\sqrt{[(E^2)_{\rm R}]^2+[(E^2)_{\rm I}]^2}$.}
\begin{eqnarray}
E_{\rm I}(T,0)={(E^2(T,0))_{\rm I}\over 2E_{\rm R}(T,0)}
={(M^2(T))_{\rm I}-\varepsilon \over 2E_{ \rm R}(T,0)}
={2M_{\rm R}(T)M_{\rm I}(T)-\varepsilon \over 2E_{\rm R}(T,0)}=0.
\end{eqnarray}
Therefore, the imaginary part of the effective mass behaves as
\begin{eqnarray}
 M_{\rm I}(T)={\varepsilon \over 2M_{\rm R}(T)},
\end{eqnarray}
which increases with decreasing $M_{\rm R}(T)$.

Using Eq.(3$\cdot$5), we have
\begin{eqnarray}
E_{\rm I}(T,q)
={2M_{\rm R}(T)M_{\rm I}(T)-\varepsilon \over 2E_{\rm R}(T,q)}=0
\end{eqnarray}
for all $q$. The real part of the energy is given by
\begin{eqnarray}
E_{\rm R}(T,q)=\sqrt{(E^2(T,q))_{\rm R}+|E^2(T,q)| \over 2}=\sqrt{q^2+(M^2(T))_{\rm R}}
\end{eqnarray}
for $q^2+(M^2(T))_{\rm R}>0$. Since $M_{\rm R}(T)>M_{\rm I}(T)$ as shown in Figs.1,2 and 5, $(M^2(T))_{\rm R}>0$ is satisfied in the considered region.

 From Eq.(3$\cdot$7) with $q=0$, the real part of the energy becomes 
\begin{eqnarray}
E_{\rm R}(T,0) =\sqrt{(M^2(T))_{\rm R}}
=\sqrt{M_{\rm R}^2(T)-\left({\varepsilon\over 2M_{\rm R}(T)}\right)^2}\simeq \sqrt{M_{\rm R}^2(T)},
\end{eqnarray}
since $M_{\rm R}^2(T)\gg \varepsilon $ in this region.
From equation (3$\cdot$8), it can be seen that as the temperature increases, the real part of the energy decreases with the real part of the effective mass.

 Above results give that the denominator of the effective propagator  is 
$$
 D(T,z,q)= z^2-q^2-M^2(T)+i\varepsilon=z^2-q^2-(M^2(T))_{\rm R}
$$
 \begin{eqnarray}
 =z^2-E_{\rm R}^2(T,q)=(z-E_{\rm R}(T,q))(z+E_{\rm R}(T,q)),
\end{eqnarray} 
since $(M^2(T))_{\rm I}=\varepsilon$. 
Therefore, the poles of the effective propagator are on the real axis.

From Fig.5, near $T_{\rm C}$, the decrease in the real part of the effective mass is understood to be a process of restoration of the chiral symmetry.
On the other hand, the imaginary part of the effective mass increases rapidly, though it has finite value.
 The decrease in the imaginary part of the effective mass  beyond the peak is also understood to be the process of restoration of the chiral symmetry. 
When $M_{\rm R}(T)$ and $M_{\rm I}(T)$ become small, the relation $E_{\rm I}(T,0)=0$ is broken. Therefore, the relation in Eq.(3$\cdot5$) also does not hold. 
In this region, $E_{\rm R}(T,0)$ approaches $\sqrt{\varepsilon/2}$, so $E_{\rm I}(T,0)$ approaches $-\sqrt{\varepsilon/2}$, which will be shown in the following consideration, \footnote{For solutions where the real and imaginary parts of the effective mass are negative, the same considerations can be made by replacing the real and imaginary parts of the effective mass with their absolute values. Therefore, the sign of the imaginary part of energy does not change.} 

 Next, form Figs.1 and 2, the effective mass vanishes in $T>T_{\rm C}$. In this region, the chiral symmetry, including the imaginary part, is restored.
However, the real and imaginary parts of the energy are constant values for $q=0$.

From $M_{\rm R}(T)=M_{\rm I}(T)=0$, we get 
\begin{eqnarray}
E_{\rm R}(T,q)=
\sqrt{(E^2(T,q))_{\rm R}+|E^2(T,q)| \over 2}=
\sqrt{q^2+\sqrt{q^4+\varepsilon^2}\over 2}
\end{eqnarray}
and 
\begin{eqnarray}
E_{\rm I}(T,q)={-\varepsilon\over 2E_{\rm R}(T,q)}
=-{\varepsilon \over 2}\sqrt{2 \over q^2+\sqrt{q^4+\varepsilon^2}}
=-\sqrt{-q^2+\sqrt{q^4+\varepsilon^2} \over 2}.
\end{eqnarray}
Therefore, we have $E_{\rm R}(T,0)=\sqrt{\varepsilon/ 2}$ and $E_{\rm I}(T,0)=-\sqrt{\varepsilon/ 2}$ at $q=0$.
 From Eqs.(3$\cdot$10) and (3$\cdot$11), we have $(E^2(T,q))_{\rm R}=E_{\rm R}^2(T,q)-E_{\rm I}^2(T,q)=q^2$ and 
$(E^2(T,q))_{\rm I}=2E_{\rm R}(T,q)E_{\rm I}(T,q)=-\varepsilon$.

Above results give that the denominator of the effective propagator  is 
\begin{eqnarray}
D(T,z,q)= 
z^2-(E^2(T,q))_{\rm R}-i(E^2(T,q))_{\rm I}
=z^2-q^2+i\varepsilon,
\end{eqnarray}
which is the denominator of the Feynman propagator for a massless fermion.

\section{Summary and Comments}

The analytical properties of fermion propagators in the non-perturbative region  have been studied using various methods, but they are not yet understood in detail.
In particular, it is necessary to clarify the presence or absence of complex poles in the fermion propagator. Regarding this problem, it is important to know the behavior of the imaginary part of the energy or the effective mass. 
Further researches using various methods are desired to investigate these issues. Moreover, there are still many unknowns about the temperature dependence of the pole position of the fermion propagator in the complex plane. 

One of useful methods for calculating the propagator in the non-perturbative region is the Schwinger-Dyson equation (SDE). 
In this paper, we extended the previously studied our method to finite temperature, which is slightly different from other methods. In our method, the mass and energy in the SDE are extended to complex values from the beginning without assuming a spectral representation, and the propagator in the self-energy  is integrated on the complex energy plane. 
 Then we examined the properties of the solutions obtained by the SDE on the complex plane. To confirm the results we obtained, we compared the calculations in two different ways and found that the two results agreed.

We explained our formulation and analysis method using a simplified four-fermion contact interaction model in the $(1+1)$ space-time dimensions, in which the effective mass does not depend on the energy and momentum.
The real and imaginary parts of the SDE were solved simultaneously by the iterative method.   In this model, the chiral symmetry for the fermion is broken below a phase transition temperature $T_{\rm C}$, which depends on the coupling constant. 

In the temperature below $T_{\rm C}$, except near $T_{\rm C}$, 
  even though the imaginary part of the effective mass has a finite value, the imaginary part of the energy vanishes.
  Therefore, the poles of the effective propagator lie on the real axis. This is a different situation in strongly coupled quantum electrodynamics (QED) and quantum chromodynamics (QCD), where the poles are not on the real axis as pointed out by the various studies mentioned in the introduction. 
    
  Moreover, as shown in Fig.1, the real part of the effective mass begins to decrease as it approaches the phase transition temperature, but Fig.2 shows that the change in the imaginary part of the effective mass has a peak near the phase transition temperature. As can be seen from Fig.5, this peak has a narrow width and a finite height.  Therefore, the real and imaginary parts of the effective mass behave differently near the phase transition temperature.
  
  At temperatures above $T_{\rm C}$, the effective mass including the imaginary part vanishes and chiral symmetry is restored. However, the real and imaginary parts of the energy have finite values.
  In this region, the poles of the effective propagator  are the same as the usual Feynman propagator for a massless fermion. 
 
 Due to the temperature dependence of the energy as described above, the poles of the effective propagator change rapidly around the phase transition temperature. It seems to be an interesting problem to investigate the temperature dependence of the poles of the effective propagator on the complex energy plane in various models.
    
Although this paper dealt with a simple model, it seems that the temperature dependence of the effective propagator on the complex energy plane in the non-perturbative region is interesting to study. 
 Examining the behavior of the effective propagator on the complex energy plane using the SDE for various theories is expected to be meaningful in obtaining some hints for understanding the properties of quantum field theory  in the non-perturbative region.


\vspace{5mm}
\begin{center}
{\large \bf Appendix A. Formulas for numerical calculations}
\end{center}

\vspace{5mm}

The SDE is given in Eq.(3$\cdot$1) as
$$ M(T)
=2\pi\lambda\int_{\delta}^{\Lambda} dq { M(T) \over  E(T,q)} \tanh\left({1 \over 2}\beta E(T,q)\right)$$
with $\beta=1/T$.

From 
 $$\zeta(T,q) = \tanh\left({1 \over 2}\beta  E(T,q)\right)={e^{\beta E(T,q)/2}-e^{-\beta E(T,q)/2} \over e^{\beta E(T,q)/2}+e^{-\beta E(T,q)/2}}
 =1-{2 \over 1+e^{\beta E(T,q)}}$$
$$ =1-2{1+e^{\beta E_{\rm R}(T,q)}[\cos(\beta E_{\rm I}(T,q))-i\sin(\beta E_{\rm I}(T,q))] \over 1+2e^{\beta E_{\rm R}(T,q)}\cos(\beta E_{\rm I}(T,q))+e^{2\beta E_{\rm R}(T,q)}},$$
 the real and imaginary parts of $\zeta(T,q)$ are given by
 $$\zeta_{\rm R}(T,q)=1-2{1+e^{\beta E_{\rm R}(T,q)}\cos(\beta E_{\rm I}(T,q)) \over 1+2e^{\beta E_{\rm R}(T,q)}\cos(\beta E_{\rm I}(T,q))+e^{2\beta E_{\rm R}(T,q)}}$$
and
 $$\zeta_{\rm I}(T,q)=2{e^{\beta E_{\rm R}(T,q)}\sin(\beta E_{\rm I}(T,q)) \over 1+2e^{\beta E_{\rm R}(T,q)}\cos(\beta E_{\rm I}(T,q))+e^{2\beta E_{\rm R}(T,q)}},$$
respectively.
 
Using $\zeta_{\rm R}(T,q)$ and $\zeta_{\rm I}(T,q)$ represented by the above formulas, the real and imaginary parts of the effective mass are written as
 $$ M_{\rm R}(T)
 =2\pi\lambda\int_{\delta}^{\Lambda} dq {[M(T)E^*(T,q)\zeta(T,q)]_{\rm R} \over |E(T,q)|^2}$$
and
$$ M_{\rm I}(T)
 =2\pi\lambda\int_{\delta}^{\Lambda} dq {[M(T)E^*(T,q)\zeta(T,q)]_{\rm I} \over |E(T,q)|^2}$$ 
with
 $$[M(T)E^*(T,q)\zeta(T,q)]_{\rm R}=[M(T)E^*(T,q)]_{\rm R}\zeta_{\rm R}(T,q)-[M(T)E^*(T,q)]_{\rm I}\zeta_{\rm I}(T,q)$$
and
 $$[M(T)E^*(T,q)\zeta(T,q)]_{\rm I}=[M(T)E^*(T,q)]_{\rm R}\zeta_{\rm I}(T,q)+[M(T)E^*(T,q)]_{\rm I}\zeta_{\rm R}(T,q).$$
Here, $[M(T)E^*(T,q)]_{\rm R}$ and $[M(T)E^*(T,q)]_{\rm I}$ are given by 
 $$[M(T)E^*(T,q)]_{\rm R}=M_{\rm R}(T)E_{\rm R}(T,q)+M_{\rm I}E_{\rm I}(T,q) $$
and
 $$[M(T)E^*(T,q)]_{\rm I}=M_{\rm I}(T)E_{\rm R}(T,q)-M_{\rm R}(T)E_{\rm I}(T,q),$$
 respectively.



\vspace{5mm}

\begin{center}
{\large \bf Appendix B. Alternative formulas for numerical calculations}
\end{center}

\vspace{5mm}

Introducing a ultraviolet cutoff $\Lambda$ and an infrared cutoff $\delta$ for the momentum $q$, we write  the SDE for the effective mass $M(T)$ in the imaginary-time formalism given by Eq.(2$\cdot$2) as
$$
  M(T)= 2\pi T\lambda \int_{-\Lambda}^{\Lambda} dq \sum_{n=-\infty}^{\infty} {M(T)\Theta(|q|-\delta) \over  \omega_n^2+q^2+M^2(T)-i\varepsilon}\equiv 2\pi T\lambda \sum_{n=-\infty}^{\infty} M(T)I_n(T),
$$
where
$$ I_n(T)=\int_{-\Lambda}^{\Lambda} dq {\Theta(|q|-\delta)  \over  \omega_n^2+q^2+M^2(T)-i\varepsilon}
\equiv \int_{\delta}^{\Lambda} dq {2 \over q^2-\chi_n^2(T)}
= \int_{\delta}^{\Lambda} dq{2 \over  (q-\chi_n(T))(q+\chi_n(T))} $$
with $\omega_n=2\pi T(n+1/2)~(n=0,\pm 1,\pm 2,\cdots)$ and 
$$ \chi_n^2(T)= -\omega_n^2-M^2(T)+i\varepsilon.$$
Here, we define $\left(\chi_n(T)\right)_{\rm R}>0$.

The real and imaginary parts of $\chi_n(T)$  are given as  
$$(\chi_n(T))_{\rm R}=\sqrt{(\chi^2_n(T))_{\rm R}+|\chi^2_n(T)| \over 2}
={1 \over \sqrt{2}}\sqrt{(\chi^2_n(T))_{\rm R}+\sqrt{[(\chi^2_n(T))_{\rm R}]^2+[(\chi^2_n(T))_{\rm I}]^2}}$$
and
$$(\chi_n(T))_{\rm I}={(\chi^2_n(T))_{\rm I}\over 2(\chi_n(T))_{\rm R}}
={(\chi^2_n(T))_{\rm I}\over \sqrt{2}|(\chi^2_n(T))_{\rm I}|}\sqrt{-(\chi^2_n(T))_{\rm R}+\sqrt{[(\chi^2_n(T))_{\rm R}]^2+[(\chi^2_n(T))_{\rm I}]^2} },$$
respectively, in which
 we define 
$ (\chi^2_n(T))_{\rm R}=-\omega_n^2-(M^2(T))_{\rm R}=-\omega_n^2-M_{\rm R}^2(T)+M_{\rm I}^2(T)$,
$(\chi^2_n(T))_{\rm I}=-(M^2(T))_{\rm I}+\varepsilon= -2M_{\rm R}(T)M_{\rm I}(T)+\varepsilon $ and $|\chi^2_n(T)|=\sqrt{[(\chi^2_n(T))_{\rm R}]^2+[(\chi^2_n(T))_{\rm I}]^2}$.

Here, we write 
$$ I_n(T)={1 \over 2\chi_n(T)}\left[{\bar I}_n^{(-)}(T)-{\bar I}_n^{(+)}(T)\right]
\equiv {2 \over 2\chi_n(T)}{\bar I}_n(T)$$
with
$$ {\bar I}_n^{(\pm)}(T)=\int_{\delta}^{\Lambda} dq{1 \over q\pm \chi_n(T)}.$$

Integrating over $q$, the real and imaginary parts of ${\bar I}_n^{(\pm)}(T)$ are given by 
$$({\bar I}_n^{(\pm)}(T))_{\rm R}={1 \over 2}\log{(\Lambda\pm (\chi_n(T))_{\rm R})^2+(\chi_n(T))_{\rm I}^2 \over (\delta\pm (\chi_n(T))_{\rm R})^2+(\chi_n(T))_{\rm I}^2}$$
and 
$$({\bar I}_n^{(\pm)}(T))_{\rm I}=\mp {(\chi_n(T))_{\rm I} \over |(\chi_n(T))_{\rm I}|}\left[\arctan{\Lambda\pm (\chi_n(T))_{\rm R}\over |(\chi_n(T))_{\rm I}|}-\arctan{\delta\pm (\chi_n(T))_{\rm R}\over |(\chi_n(T))_{\rm I}|}\right],$$
respectively.

From $ I_n(T)= {\bar I}_n(T)/\chi_n(T)$, the real and imaginary parts of $ I_n(T)$ are given by 
$$ (I_n(T))_{\rm R}={(\chi_n(T))_{\rm R}({\bar I}_n(T))_{\rm R}+(\chi_n(T))_{\rm I}({\bar I}_n(T))_{\rm I} \over |\chi_n(T)|^2}$$
and 
$$ (I_n(T))_{\rm I}={(\chi_n(T))_{\rm R}({\bar I}_n(T))_{\rm I}-(\chi_n(T))_{\rm I}({\bar I}_n(T))_{\rm R} \over |\chi_n(T)|^2},$$
respectively.

Therefore, the SDEs for the real and imaginary parts of the effective mass are given by
$$
  M_{\rm R}(T)
  =2\pi T\lambda \sum_{n=-\infty}^{\infty} [M(T)I_n(T)]_{\rm R}$$
  and
$$M_{\rm I}(T)
  =2\pi T\lambda \sum_{n=-\infty}^{\infty} [M(T)I_n(T)]_{\rm I}$$
  with
$$[M(T)I_n(T)]_{\rm R}=M_{\rm R}(T)(I_n(T))_{\rm R}-M_{\rm I}(T)(I_n(T))_{\rm I}$$
and
$$[M(T)I_n(T)]_{\rm I}=M_{\rm R}(T)(I_n(T))_{\rm I}+M_{\rm I}(T)(I_n(T))_{\rm R},$$
respectively. The numerical results using the above formulas reproduce the results presented in Figs.1 and 2.

Below, we will roughly estimate the relation between the coupling constant $\lambda$ and the phase transition temperature $T_{\rm C}$.
Since $M_{\rm R}(T)\gg M_{\rm I}(T)$ from numerical results presented in Figs.1, 2 and 5 for $T < T_{\rm C}$, we ignore the imaginary part $M_{\rm I}(T)$ for simplicity.
 Additionally, we set $\varepsilon=\delta=0$. Then we have $(\chi_n(T))_{\rm R}=0$ and $(\chi_n(T))_{\rm I}=\sqrt{\omega_n^2+M_{\rm R}^2(T) }$. Using these results, we get $ ({\bar I}_n(T))_{\rm R}=0$ and
$$ ({\bar I}_n(T))_{\rm I}=2s_n(T)\arctan{\Lambda\over |(\chi_n(T))_{\rm I}|}$$
with $s_n(T)=(\chi_n(T))_{\rm I}/|(\chi_n(T))_{\rm I}|=(\chi_n^2(T))_{\rm I}/|(\chi_n^2(T))_{\rm I}|$. Therefore, we have
$$ (I_n(T))_{\rm R}=2{1 \over |(\chi_n(T))_{\rm I}|}\arctan{\Lambda\over |(\chi_n(T))_{\rm I}|}$$
and $(I_n(T))_{\rm I}=0$, respectively.

Using these results, the SDEs for the effective mass become 
$$ M_{\rm R}(T)=2\pi T\lambda \sum_{n=-\infty}^{\infty} M_{\rm R}(T)(I_n(T))_{\rm R}$$
and $M_{\rm I}(T)=0$. From the real part of the SDEs,
$$1=2\pi T\lambda \sum_{n=-\infty}^{\infty} (I_n(T))_{\rm R}
=4\pi (T/\Lambda)\lambda \sum_{n=-\infty}^{\infty}{\Lambda \over |(\chi_n(T))_{\rm I}|}\arctan{\Lambda\over |(\chi_n(T))_{\rm I}|}$$
should be satisfied for $M_{\rm R}(T)\neq 0$. 

Since, if $\lambda$ is not too small, $\Lambda/|(\chi_n(T))_{\rm I}|\ll 1$ and $\omega_n^2 \gg M_{\rm R}^2(T)$  are satisfied near $T=T_{\rm C}$,\footnote{For $\lambda=0.3$ and $T\simeq T_{\rm C}\simeq 0.9$, each $\Lambda/|(\chi_n(T))_{\rm I}|$ is less than $5\times 10^{-3}$. For larger $\lambda$, $T_{\rm C}$ becomes larger, so our approximation may be no worse.}
the above equation can be approximated to
$$ {1 \over 4\pi\lambda T/\Lambda}= \sum_{n=-\infty}^{\infty}{\Lambda \over |(\chi_n(T))_{\rm I}|}\arctan{\Lambda\over |(\chi_n(T))_{\rm I}|}
\simeq  \sum_{n=-\infty}^{\infty}\left({\Lambda \over |(\chi_n(T))_{\rm I}|}\right)^2  $$
$$=\sum_{n=-\infty}^{\infty}{\Lambda^2 \over \omega_n^2+M_{\rm R}^2(T)}
\simeq \Lambda^2\sum_{n=-\infty}^{\infty}{1 \over \omega_n^2}\left(1-{M_{\rm R}^2(T)\over \omega_n^2}\right).$$
Taking the sum of $n$, we have
$${1 \over 4\pi\lambda T/\Lambda}\simeq {6 \zeta(2)\over (2\pi T/\Lambda)^2}
-{M_{\rm R}^2(T)\over \Lambda^2}{30 \zeta(4)\over (2\pi T/\Lambda)^4},$$
which gives
$${M_{\rm R}^2(T)\over \Lambda^2}\simeq {(2\pi T/\Lambda)^2 \over 30\zeta(4)}\left[6 \zeta(2)-{\pi(T/\Lambda) \over\lambda }\right].$$
Here, $\zeta(2)$ and $\zeta(4)$ are the Riemann zeta functions.
If $M_{\rm R}^2(T)<0$, there is no nontrivial solution.
From this consideration, the phase transition temperature $T_{\rm C}$ may be  approximately given by
$$ {T_{\rm C} \over \Lambda}\simeq {6 \zeta(2) \over \pi}\lambda 
=\pi\lambda$$
with $\zeta(2)=\pi^2/6$, which gives, for $\lambda=0.3,0.6,0.9$, the phase transition temperature $T_{\rm C}/\Lambda\simeq 0.94, 1.88, 2.83$, respectively.

\newpage

\vspace{5mm}

\begin{flushleft} 
{\bf References }
\end{flushleft}

\vspace{2mm}


\begin{description}
\item{[1]}  F. J. Dyson, Phys.Rev.{\bf 75}, 1736(1949).
\item{[2]}  J. S. Schwinger, Proc.Nat.Acad.Sci.{\bf 37}, 452(1951).
\item{[3]} R. Alkofer and L. von Smekal, Phys.Rep.{\bf 353}, 281(2001) [arXiv:hep-ph/0007355 ].
\item{[4]} C. S. Fischer, J. Phys.G{\bf 32}, R253(2006)  [arXiv:hep-ph/0605173].
\item{[5]} M. Q. Huber, Phys.Rept.{\bf 879}, 1(2020) [arXiv:1808.05227 [hep-ph]].
\item{[6]} A. Cucchieri, D. Dudal, T. Mendes, and N. Vandersickel, Phys.Rev.D{\bf 85}, 094513(2012) [arXiv:1111.2327 [hep-lat]]. 
\item{[7]} F. Falcao, O. Oliveira, and P. J. Silva, Phys.Rev.D{\bf 102}. 114518 (2020) [arXiv:2008.02614 [hep-lat]].
\item{[8]} D. Boito, A. Cucchieri, C. Y. London, and T. Mendes, JHEP{\bf 02}, 144(2023) [arXiv:2210.10490 [hep-lat]].
\item{[9]} A. Cucchieri, T. Mendes, and A. R. Taurines, Phys.Rev.D{\bf 71}, 051902 (2005) [arXiv:hep-lat/0406020].
\item{[10]} J. M. Cornwall, Mod.Phys.Lett.A{\bf 28}, 1330035 (2013) [arXiv:1310.7897 [hep-ph]].
\item{[11]} D. Dudal, O. Oliveira, and P. J. Silva, Phys.Rev.D{\bf 89}, 014010(2014) [arXiv:1310.4069 [hep-lat]].
\item{[12]} V. N. Gribov, Nucl.Phys.B{\bf 139}, 1 (1978).
\item{[13]} D. Zwanziger, Nucl.Phys.B{\bf 364}, 127 (1991).
\item{[14]} D. Dudal, S. P. Sorella, N. Vandersickel, and H. Verschelde, Phys.Rev.D{\bf 77}, 071501(2008) [arXiv:0711.4496 [hep-th]].
\item{[15]} D. Dudal, J. A. Gracey, S. P. Sorella, N. Vandersickel, and H. Verschelde, Phys.Rev.D{\bf 78}, 065047(2008) [arXiv:0806.4348 [hep-th]].
\item{[16]} N. Vandersickel and D. Zwanziger, Phys.Rept.{\bf 520}, 175(2012) [arXiv:1202.1491 [hep-th]].
\item{[17]} G. P. de Brito, P. de Fabritiis, and A. D. Pereira, Phys.Rev.D{\bf 107}, 114006(2023) [arXiv:2302.04827 [hep-th]].
\item{[18]}  S. Strauss, C. S. Fischer, and C. Kellermann, Phys.Rev.Lett.{\bf 109}, 252001(2012) [arXiv:1208:6239 [hep-ph]].
\item{[19]} Y. Hayashi and K.-I. Kondo, Phys.Rev.{\bf D103}, L111504 (2021) [arXiv:2103.14322].
\item{[20]} Y. Hayashi and K.-I. Kondo, Phys.Rev.{\bf D104}, 074024(2021) [arXiv:2105.07487]. 
\item{[21]} R. Alkofer, W. Detmold, C. S. Fischer, and P. Maris, Phys.Rev.D{\bf 70}, 014014(2004) [hep-ph/0309077].
\item{[22]} C. S. Fischer and M. Q. Huber, Phys.Rev.{\bf D102}, 094005(2020) [arXiv:2007.11505].
\item{[23]} P. Maris, Phys.Rev.{\bf D50}, 4189(1994).
\item{[24]} P. Maris, Phys.Rev.{\bf D52}, 6087(1995) [arXiv:hep-ph/9508323].
\item{[25]} P. Bicudo, Phys. Rev. D{\bf 69}, 074003 (2004) [arXiv:hep-ph/0312373].
\item{[26]} V. Sauli and Z. Batiz, J.Phys.G {\bf 36}, 035002(2009) [arXiv:0806.2817 [hep-ph]].
\item{[27]} V. Sauli and Z. Batiz, Few Body Syst.{\bf 48}, 41(2010) [arXiv:1006.0612 [hep-ph]].
\item{[28]} V. Sauli, J. Adam, Jr., and P. Bicudo, Phys.Rev.D{\bf 75}, 087701(2007) [arXiv:hep-ph/0607196].
\item{[29]} S. Jia and M. R. Pennington, Phys.Rev.D{\bf 96}, 036021(2017) [arXiv:1705.04523 [nucl-th]].
\item{[30]} V. Sauli, Few Body Syst.{\bf 61}, 3,23(2020) [arXiv:1809.07644 [hep-ph]].
\item{[31]} C. Mezrag and  G. Salm\'e, Eur.Phys.J.C{\bf 81}, 1,34(2021) [arXiv: 2006.15947 [hep-ph]].
\item{[32]} D. C. Duarte, T. Frederico, W.de Paula, and E. Ydrefors, Phys.Rev.D{\bf 105}, 114055(2022) [arXiv:2204.08091 [hep-ph]].
\item{[33]} V. Sauli, Phys.Rev.D{\bf 106}, 094022(2022) [arXiv:2011.00536 [hep-lat]].
\item{[34]} A. F. Falcao and O. Oliveira, Phys.Rev.D{\bf 106}, 114022(2022) [arXiv:2209.14815 [hep-lat]].
\item{[35]} J. Horak, J. M. Pawlowski, and N. Wink, SciPost Phys.{\bf 15}, 4,149(2023) [arXiv:2210.07597 [hep-ph]].
\item{[36]} J. A. Mueller, C. S. Fischer, and D. Nickel, Eur.Phys.J.C{\bf 70}, 1037(2010) [arXiv:1009.3762 [hep-ph]].
\item{[37]} S.-X. Qin, L. Chang, Y.-X. Liu, and C. D. Roberts, Phys.Rev.D{\bf 84}, 014017(2011) [arXiv:1010.4231 [nucl-th]].
\item{[38]} S. Schlichting, D. Smith, and L. von Smekal, Nucl.Phys.B{\bf 950}, 114868(2020) [arXiv:1908.00912 [hep-lat]].

\newpage

\item{[39]} D. Schweitzer, S. Schlichting, and L. von Smekal, Nucl.Phys.B{\bf 960}, 115165(2020) [arXiv:2007.03374 [hep-lat]].
\item{[40]} H. Tanaka and S. Sasagawa, Prog.Theor.Exp.Phys.{\bf 2023}, 7,073B02(2023) [arXiv:2301.11793 [hep-th]]. 
\item{[41]} D. J. Gross and A. Neveu, Phys.Rev.{\bf D10}, 3235(1974).
\item{[42]} L. Jacob, Phys.Rev.{\bf D10}, 3956(1974). 
\item{[43]} N. Dorey and N. E. Mavromatos, Phys.Lett.{\bf B266}, 163(1991).
\item{[44]} K. Fukazawa, T. Inagaki, S. Mukaigawa, and T. Muta, Prog.Theor.Phys.{\bf 105},979(2001) [arXiv:hep-ph/9910305].
\item{[45]} B. J. Harrington and A. Yildiz, Phys.Rev.{\bf D11}, 779(1975).
\item{[46]} S. Ma and R.Rajaraman, Phys.Rev.{\bf D11}, 1499(1975).
\item{[47]} S.-Z. Huang and M. Lissia, Phys.Rev.{\bf D53}, 7270(1996) [arXiv:hep-ph/9509360 [hep-ph]].
\item{[48]} B.-R. Zhou, Commun.Theor.Phys.{\bf 39}, 663(2003) [arXiv:hep-ph/0212193].
\item{[49]} J.-L. Kneur, M. B. Pinto, and R. O. Ramos, Braz.J.Phys.{\bf 37}, 258(2007) [arXiv:0704.2843 [hep-ph]].
\item{[50]} M. Le Bellac, Thermal Field Theory, Cambridge Monographs on mathematical physics, Cambridge university press (1996).
\end{description}

\end{document}